\RequirePackage[hyphens]{url}
\documentclass{acm_proc_article-sp}  
\pdfoutput=1

\begin{document} \sloppy

\title{A survey of systems for massive stream analytics}
%
%
%
%
%

\numberofauthors{3} 
%
\author{
%
%
Maninder Pal Singh, Mohammad A. Hoque, Sasu Tarkoma\\  
       \affaddr{University of Helsinki}\\
       \affaddr{Department of Computer Science}\\
       \affaddr{FI-00014, Finland}\\        
       \email{firstname.lastname@cs.helsinki.fi}
}

\date{29 September 2014}

\maketitle
\begin{abstract}
The immense growth of data demands switching from traditional data processing solutions to systems, which can process a continuous stream of real time data. Various applications employ stream processing systems to provide solutions to emerging Big Data problems.  Open-source solutions such as Storm, Spark Streaming, and S4 are the attempts to answer key stream processing questions.  The recent introduction of real time stream processing commercial solutions such as Amazon Kinesis, IBM Infosphere Stream reflect industry requirements.  The system and application related challenges to handle massive stream of real time data analytics are an active field of research.

In this paper, we present a comparative analysis of the existing state-of-the-art stream processing solutions. We also include various application domains, which are transforming their business model to benefit from these large scale stream processing systems. 

\end{abstract}




\section{Introduction}

The growth of massive data domains such as social networks, high frequency trading, online gaming, advertisement, DNA sequencing are beyond the reach of traditional data processing systems. Companies are focusing towards real time data-based products for their consumers. For example, Supercell\footnote{http://www.supercell.net/} provides online games - Clash of Clans\footnote{http://www.supercell.net/games/view/clash-of-clans} and Boom Beach\footnote{http://www.supercell.net/games/view/boom-beach}. These are online combat strategy games,  which can be played on hand-held devices, such as tablets and smartphones. Supercell is using Amazon Kinesis \cite{Amazon:Kinesis:Main:Web} for real time processing of data streams generated from various devices. Amazon Kinesis helps in real time analysis of game insight from the  data originated from hundreds of users and the game engine servers \cite{AWS:Supercell}. The timely insight data helps in business analytics and to improve the game experience of the players \cite{AWS:Supercell}.  

The stream processing concept has evolved from stream computing paradigm, which involves continuous query and real time analytics on massive stream of data.  There are a number of solutions, which address real time stream processing.  S4 \cite{Neumeyer:2010:SDS:1933306.1934385}, Storm \cite{ApacheStorm} and Spark Streaming \cite{SparkStreaming:Online} are examples of existing open-source solutions.   Commercial solutions such as Amazon Kinesis \cite{Amazon:Kinesis:Main:Web}, IBM Infosphere stream \cite{IBM:Infosphere:Stream:WhitePaper} are also working in the same direction.

The paper focuses on discussion and comparative analysis of the existing state-of-the-art stream processing solutions. The rest of the paper is organized as follows. Section two discusses massive data streaming concepts. Section three looks into various streaming solutions. Section four discusses the architecture perspective of stream processing solutions and explores programming model, latency, data pipelines, fault tolerance, and data. Section five presents emerging use cases of stream processing solutions.  Section six explores the challenges of stream processing solutions and applications. Finally, the conclusions are summarized in section seven.


\section{Streaming Concepts}

Streaming data is fundamentally different from traditional data handling patterns and comes with its own set of challenges and requirements.  It requires in-stream processing to have a low latency.  The system should be scalable with self load balancing capability and should have high availability. It may require some persistent storage for short period of time. Real time streaming data has all the well-known attributes of Big Data, such as volume, variety, velocity and veracity.

Stream processing requires handling of varying rate of streaming data, the incoming streaming data might involve missing data or delays. The processing includes on-the-fly decision and provision for handling such out of order streaming data. The streaming data can be time-stamped on arrival or subject to discard depending upon sensitiveness of data. The time sensitive operations require time-out of blocking computations. The time intensive operations require careful handling of stream linked to the system and resources binding to the computation. Real time streaming also requires deterministic processing as time-order guarantee is subject to different conditions. The system demands mining around processed streaming data and data stored across persistent storage. Although the persistent storage adds additional latency, the integration is required to provide business analytic around data. Persistent storage for long time of streaming data involves its own set of challenges. The streaming data require extensible framework for querying and processing to conclude desired results. The operators involving stream data require understanding of streaming data context. The variety of data in the structured, unstructured or semi-structured form requires adaptability in real time processing.  

The attributes such as data integrity and data availability are the integral parts of data engines. There is a change in paradigm, which involves distributed processing solution providers to run on low cost commodity hardware clusters. The system demands scalability and transparent load-balancing for such high volume of data.  The data engine should be adaptive, extensible to add easy to program modules, and capable to process high-volume of streaming data with low latency.
 
\section{Streaming Solutions}

There are few solutions available for real time massive streaming data processing. The available solutions can be classified into open source contributions and commercial solutions. Table \ref{table:kysymys} refers to various available open-source solutions. Table \ref{table:propStream} refers to the list of various commercial solutions. 


\def\arraystretch{1.2}%
\begin{table}
\centering
\caption{Open Source Streaming Solutions}
\begin{tabular}{|l|l|l|} \hline Solution&Type&Developed By\\ \hline

Storm & Streaming & BackType\\ \hline
Spark Streaming & Batch \& Streaming & Berkeley AMPLab  \\ \hline
S4 & Streaming & Yahoo \\  
\hline\end{tabular}
\label{table:kysymys}
\end{table}

\def\arraystretch{1.2}%
\begin{table}
\centering
\caption{Commercial Streaming Solutions}
\begin{tabular}{|l|l|} \hline Solution&Developed By\\ \hline

MillWheel &  Google  \\ \hline
Amazon Kinesis &  Amazon\\ \hline
Infosphere &  IBM  \\ \hline
\end{tabular}
\label{table:propStream}
\end{table}

\subsection{Open Source Streaming Solutions}

\textbf{Storm} is a distributed stream processing framework, developed in Clojure and built upon model of task parallel computation  \cite{ApacheStorm}. It provides an adapter to write applications  virtually in any language. Storm is optimized for low-latency processing and uses ZeroMQ\footnote{http://zeromq.org/} for message passing, which makes its architecture to provide a guaranteed message processing \cite{ApacheStorm}. It attempts to process each record at least once and if a record is not yet processed by a node, it replays the records. In addition, It provides fair fault detection and process management. On discovery of the failure of a task, messages are automatically reassigned by quickly restarting the processing. For optimal resource handling, the processes in Storm are managed by some supervisors. 

\textbf{Spark Streaming} is an extension of the core Spark which is an in-memory distributed data analysis platform \cite{SparkStreaming:Online}. Spark is built upon the model of data parallel computation. It provides reliable processing of live streaming data. Spark streaming transforms streaming computation into a series of deterministic micro-batch computations, which are then executed using Spark's distributed processing framework.
 
\textbf{S4} (Simple Scalable Streaming System) is a general purpose distributed and scalable streaming platform that allows the processing of continuous unbounded streams of data. Its processing model is inspired by MapReduce, which uses key based programming model \cite{gdfm:S4vsStorm:Online}. The computation is performed by processing elements and messages are transmitted between them in the form of data events \cite{Neumeyer:2010:SDS:1933306.1934385}.

\subsection{Commercial Streaming Solutions}

\def\arraystretch{1.2}%
\begin{table*}[!ht]
\begin{minipage}{\textwidth} 
\centering
\caption{Attributes based Streaming solution classification}
\begin{tabular}{| L{4cm} | L{4cm} | L{4cm} | L{4cm} |} \hline Attributes& Storm & Spark Streaming & S4\\ \hline

Framework & Stream Processing + Micro-Batching using Trident & Micro-Batching with Batch Processing using Core Spark & Actor Programming Model \\ \hline  

Implemented in & Clojure & Scala & Java \\  \hline 

Application Language & Java, Clojure, Scala, Python, Ruby  & Java, Scala & Java, Python, C++ \\ \hline

Stream Primitive & Tuples & DStream 	  & Events \\ \hline

Stream Source & Spouts & Network, HDFS & Network \\ \hline

Computation or Transformation  & Bolts & Transformation, Window Operations & Processing Element \\ \hline

Persistence Entity & Bolts & foreach RDD & Control Messages \\ \hline

Reliable  Execution & At least once & Exactly once & -- \\ \hline 

Fault Tolerance & Tuples are replayed, Guaranteed delivery & Tiny bits loss possible, Require HDFS for fully fault tolerant & New Node begin from snapshot \\  \hline

Latency & Sub-Second & Few Seconds & Few Seconds\footnote{Assuming low latency as few seconds \cite{Neumeyer:2010:SDS:1933306.1934385}} \\ \hline

Developed By & Conceived by BackType/ Twitter, Now Apache incubation project & Conceived by AmpLab Berkely, Now Apache incubation project & Initially conceived by Yahoo!, Now Apache incubation project \\

\hline\end{tabular}
\label{table:tab3}
\end{minipage}
\end{table*}

\textbf{Google\footnote{https://www.google.com/} MillWheel} is a framework for low-latency  streaming data processing applications \cite{41378}. It is also inspired by MapReduce programming model and allows users to write application logic in a directed computer graph using custom topology \cite{41378}. Records in a Google MillWheel are delivered continuously along the edges in a graph \cite{41378}. It provides fault tolerance and guaranteed delivery of records to the users. 
 
\textbf{Amazon\footnote{http://www.amazon.com/} Kinesis} is another service to process real time massive data from streams \cite{a2zKinesis}. It is a recent solution, which was introduced in late 2013. Kinesis adapts to streaming data and do load-balancing by auto-scaling. Fault tolerance is provided by check-pointing to replay the data records. It comes with a Kinesis client library that requires users to create "Producer" and "Worker" in an application. The Producer accepts data from a stream source and converts it into a Kinesis stream. Kinesis stream consists of data records organized into tuples. The Worker acts as a client application, which accepts Kinesis stream and performs required processing on stream. The worker can be invoked on a stream of data to obtain required results. The processed data is available only for 24 hours, which requires a user to link storage solution for future processing.

\textbf{IBM\footnote{http://www.ibm.com/} InfoSphere Streams} (Streams) is a high-performance stream processing system \cite{ballard2010ibm}. It is used for large scale continuous real time in-stream data processing \cite{ballard2010ibm}. InfoSphere does not follow specific programming model. The Stream Processing Language (SPL) has been used for developing streaming data processing applications. SPL is a declarative programming language \cite{ballard2010ibm}. SPL allows users to create complex applications without focusing on intricacies of distributed execution \cite{ballard2010ibm}. Users can control operator and its execution using C++ or Java.  

InfoSphere includes various management services, which together lay the foundation of distributed execution.  An application accepts jobs and performs concurrent processing.  A job consists of one or more Processing Elements (PEs) \cite{ballard2010ibm}.  Messaging in the system is performed using Low Latency Messaging (LLM) mechanism to optimize application execution.  IBM Infosphere Streams is actively used in diverse domains such as transportation \cite{Biem:2010:IIS:1807167.1807291}, DNA sequencing \cite{Kienzler:2011:LDS:2238436.2238494}, radio astronomy \cite{5495521}, weather forecasting \cite{daldorff2009novel}, stock market trading \cite{4812538}, and  telecommunications \cite{IBM:Infosphere:App:Telecomm}.

\section{Architecture Analysis} 

The streaming data analytics concept has been divided into micro-batch and non-batch processing techniques. Spark Streaming solution provides micro-batching of an unbounded stream. It incorporates stream processing via short interval of batches and provides linear streaming solution, which is suitable for existing batch processing infrastructure. Storm and S4 both adopted non-batch processing of streaming data. Storm also provides micro-batch processing using Trident APIs. On the other hand,  Apache S4  is entirely focused on real time stream processing and does not support micro-batch processing. 


The attribute based comparison between Storm, Spark Streaming, and S4 is presented in Table \ref{table:tab3}.  The table highlights different aspects of these solutions, which can be compared in context of processing model, data pipeline, latency,  fault tolerance, and data guarantees.

\subsection{Processing Model \& Latency}

Storm does not mandate any specific programming model. It adopted both stream and complex event processing \cite{stromjens}. It follows Directed acyclic graph (DAG) as a processing model. DAG is a directed graph with no directed cycles. Storm provides topologies that operate on the streaming data. A topology is a job and is represented as DAG. The vertex in a topology represents a worker and edges represent the data flow between the worker instances. Workers are classified as spouts and bolts. Therefore, as topologies are arranged in a  DAG, the data flows from spout to bolt and reverse flow is not possible. A spout works as an input source for the topology. Since, incoming events are processed as one record at a time, Storm has sub-second latency.    

Spark streaming follows a micro-batching programming model. It combines streaming model with batch processing model. Before processing arrived data, Spark streaming batches up events within a short time frame. The batch processing of smallest units in Spark streaming leads to few second latency.  

S4 implements the Actors programming paradigm \cite{gdfm:S4vsStorm:Online}. Processing elements are defined by the user. The messages as data events are transmitted between the processing elements \cite{Neumeyer:2010:SDS:1933306.1934385}. The triggered events are consumed by the S4 processing elements. Processing Elements interact with each other either as an event generator or consumer. S4 is inspired by the  MapReduce model.

\subsection{Data pipeline}

Storm employs pull model where events from sources are pulled by each bolt. The loss of events is possible only at ingestion time. Spouts are responsible for maintaining the event rate.  

On the other hand, Spark follows micro-batching processing model where  data streams are divided into  small batches and considered as Resilient Distributed Dataset (RDD). RDD is a distributed memory abstraction that allows in-memory computations on large clusters in a fault-tolerant manner \cite{Zaharia:2012:RDD:2228298.2228301}. RDD is the smallest processing unit and the results of RDD operations are returned as batches.

Finally, S4 is based on push model. The data events are pushed to the appropriate Processing Elements. There is a possibility of data loss in case of choking of receiver buffer.




\subsection{Fault Tolerance \& Data Guarantees}

As Storm processing model is based on a record, the state of  each record requires to be tracked as arrived in DAG nodes. Storm only guarantees processing of a record at least once. In case of failure, the records can be replayed by spout. It is quite possible to have duplicates or twice updates to the mutable state of record and the events can be lost due to various reasons. Therefore, the state recovery is important from system perspective. State recovery is also one of the required attributes  for long running operations. Storm does not provide state recovery but provides guaranteed delivery and processing of data.

Spark Streaming and Storm both provide fault tolerance and data guarantees. Stateful computation is better supported in Spark Streaming. Spark Streaming guarantees that batch level processing will be executed in an exactly once manner. In case of a node failure, Spark Streaming allows rebuilding a dataset in a node.

S4 provides state recovery via uncoordinated check pointing \cite{Chauhan:2012:PEY:2415755.2415892}. In case of failure or crash, the other nodes begin operation with recent snapshot of its state. The data events to the Processing Elements may be sent with or without guaranteed delivery. S4 also provides guaranteed delivery of control messages.


\section{Applications}

The rise of various solutions to process real time continuous stream of data reflects the trends and interest of in massive streaming data analytics. The stream processing systems are adopted by variety of applications from social media to sensing devices to astronomical telescope. An overview of such applications is provided below.

\textbf{Finance services} are based on high frequency real time trading and investment information. Most of the transactions are performed using credit cards by the customers. Banking institutions have  to take preventive measures to detect any fraud activities with credit cards \cite{Storm:FinnServ:CC:Fraud:Detection}.  For that purpose, banking sector monitors and processes multiple streams of transaction every day. The real time monitoring of transactions prevents the  likelihood of miscellaneous activities.  Therefore,  real time stream processing systems play  important roles in decision making for trading and investment.

\textbf{Medical hospitals} are also involved in using distributed stream processing for health monitoring objectives. The streams of measurement data generated from various medical instruments are processed and analyzed for proactive health diagnosis. The real time stream based monitoring tool assists doctors for diagnosis and relieves workload \cite{5431948}. 


\textbf{Smart cities} \cite{hernandez2011smart} explore urban planning to incubate human adaptive environment. The real time data from different domains is analyzed for city planning and human mobility \cite{5512826}. The urban data from cities are explored to assist government in dynamic decision making process \cite{schaffers2011smart}.  These distributed real time streams of data can be used for optimization of public transportation. It also allows people to avoid traffic congestion across different routes within a city. The urban data can also be used for constant weather  and air content monitoring.




\textbf{Radio Astronomy} involves continuous stream of data from radio telescopes. The telemetry communication process collects continuous stream of data remotely using various radio elements such as antennas, beam formers. These imaging signals are synthesized and processed real time. The final accumulated outcome is stored in a system. There are number of projects which are utilizing streaming solutions such radio astronomy group of Uppsala University and the LOFAR Outrigger In Scandinavia (LOIS) project \cite{5495521} \cite{IBM:Infosphere:Stream:WhitePaper}.  

\textbf{DNA sequence analysis} requires large-scale data set processing along with incremental computation and parallel processing while handling linear scalability. The Next-Generation Sequencing (NGS) methods benefit  from streaming data analysis in a scalable and cost-efficient way. The stream computing provides promising solution for large scale data-intensive computations in domain such as bio-informatics. The stream-based data management solution for large-scale DNA sequence analysis is explored using IBM Infosphere Streams computing platform \cite{Kienzler:2011:LDS:2238436.2238494}.

There are endless possibilities to utilize real time streaming data. Various applications such as the  personalization of web page by Yahoo!\footnote{https://www.yahoo.com/}, pay-as-you-drive insurance model, recommendation system, weather forecasting, energy trading services are the emerging domains. They are are transforming their business models to gain benefits from the analytics on real time data streams. With the major developments in Internet of Things, distributed real time stream processing  and analysis soon will be part of life.

\def\arraystretch{1.2}%
\begin{table*}[!ht]
\begin{minipage}{\textwidth} 
\centering
\caption{Applications using streaming solution in real time environment}
\begin{tabular}{| L{3cm} | L{3cm} | L{10.5cm} |} 
\hline 

	Applications	& 
    Applied Streaming Solution	& Challenges \\ \hline

	Online Gaming (esp. Supercell) \cite{AWS:Supercell} & 
    Amazon Kinesis &  
	 
    \nextitem real time data streams  originated from multiple players
    \nextitem continuous query on data streams to improve player experience
    \nextitem real time player sessions to provide real time experience
    \nextitem business analytics on real time insight of game data
    
    \\ \hline

	Medical Hospitals \cite{5431948} & 
    IBM Infosphere Stream &  
	 
    \nextitem privacy-protected real time stream monitoring from medical devices
    \nextitem analysis of data streams to explore correlation in patient diseases
    \nextitem predictive proactive medical alerts from real time data streams
    \nextitem handling multiple data streams on large scale from multiple patients 
    
    \\ \hline
    
    Radio Astronomy Imaging \cite{5495521} & 
    IBM Infosphere Stream &  
	 
    \nextitem large volume of imaging data from multiple channels
    \nextitem handling of high incoming data rate 
    \nextitem real time image synthesis for analysis
    \nextitem storage limitation as all raw data is not useful    
    
    \\ \hline
    
    DNA Sequencing \cite{Kienzler:2011:LDS:2238436.2238494} & 
    IBM Infosphere Stream &  
	 
    \nextitem large volume of genetic data
    \nextitem large-scale DNA sequence analysis
    \nextitem high latency and significant processing time
    \nextitem incremental and parallel processing    
    
    \\ \hline
    
    Smart Cities \cite{Biem:2010:IIS:1807167.1807291} & 
    IBM Infosphere Stream &  
	 
    \nextitem large volume of raw data from various source in cities
    \nextitem data disparity due to unstructured and unrelated raw data 
    \nextitem modeling of heterogeneous data and real time data analogy 
    
    \\ \hline 
    
    Finance Services \cite{Storm:FinnServ:CC:Fraud:Detection} \cite{4812538} & 
    Storm, IBM Infosphere Stream &  
	 
    \nextitem real time decision on investing and trading
    \nextitem analytics around real data stream and previous stored market data
    \nextitem monitoring of millions of high frequency transactions
    \nextitem sub-second latency
    
    \\ \hline
  
\end{tabular}
\label{table:tab4}
\end{minipage}
\end{table*}

\section{Challenges}



The stream processing solutions are designed to solve emerging Big Data trends. The solutions and applications incorporate their own set of challenges, which should be addressed before designing any solution.  The challenges require elaborate reasoning and inspection of application requirements to create an optimal solution. However, the challenges can be classified into application challenges and system level challenges. 

\subsection{Application Challenges}
 
In the earlier section we mentioned a number of domains which incorporated stream processing.  Each application is having its own set of requirements, which provide uniqueness to them. Table \ref{table:tab4} provides an overview of application challenges for domains such as radio astronomy imaging, smart cities, online gaming, medical hospitals, financial services including data handling challenges.  The high volume of data leads to high latency in DNA sequencing. The modeling of heterogeneous data and real time data analogy is a challenge for smart cities. The real time analysis of business analytics data is an important requirement for financial services.  The adaptive real time experience for players in online gaming requires continuous query on real time data.  The solutions require low latency for these domains to adapt with real time stream of massive data.

\subsection{System Challenges}

The stream processing system encounters following challenges which can be broadly categorized into four categories.

\textbf{Data Acquisition}: It is challenging to handle massive stream of continuous data.  The system requires to adapt with the velocity of incoming data. The variety of incoming data described as structured or unstructured data. The structured data acts as an optimal input for stream processing systems, whereas the unstructured data requires data pre-processing which involves filtering, extraction and organization into structured format. The latency of the stream processing system varies with structured and unstructured data. The correct representation of data and data acquisition strategies depend on the application built on the top of stream processing systems.
  
\textbf{Data Handling}: Another challenge is to properly handle large volume of data. The stream processing application requires  analyzing the sensitivity of data, which need to store into persistent storage. Some applications only require storing the cumulative processed results while other applications require  storing filtered and structurally organized processed data for later usage and analysis. The data handling and persistent storage of data format varies with the application requirement. It needs to be properly assessed by stream processing systems.
  
\textbf{Data Modeling}: The stream processing systems require in-stream processing capabilities to have a low latency. Considering the volume, variety, velocity and veracity of data, the stream processing system requires predictive models and efficient algorithms to extract application linked to important events from massive data streams. It also requires data models to perform comprehensive analysis by combining all available data.
  
\textbf{Data Mining}: The stream computing involves computational analysis and analytics around it. The stream processing requires new computational tools which can analyze heterogeneous data sets into appropriate results. It involves data analytics and data visualization of massive data sets. The traditional mining approaches need to adapt as per in-stream processing to provide dynamic results.

\section{Conclusions}


In the last decade, significant research has been performed to create a system that can handle Big Data. The MapReduce paradigm is able to offer a solution for Big Data and many solutions revolves around it. A solution based on MapReduce is suitable for many problems but not appropriate for many others. Previous research has been paired to find solutions which would be optimal for emerging Big Data trends. The stream computing paradigm appears to be a solution to the emerging Big Data trends. 

The research community is primarily focused on development of solution or mining of large data sets. The research on providing solution is divided into the selection of the programming model or data model for a system. The selection of processing model for a system varies from batch processing to micro-batch processing. Considering the availability of MapReduce as successful paradigm, many solutions for streaming are influenced by this paradigm. Some solutions also explore the Actor model to have stream processing solution. Solutions such as Storm provide a sub-second latency  and S4 does not provide persistent state and complete fault tolerance.  Spark Streaming has mixed processing model and exactly once mechanism for record delivery which might affect processing.  

A fundamental set of questions exists, which should be addressed before selecting any programming model or data model for stream processing. The design choices and challenges affect system latency and throughput. The challenges linked to applications and processing system require elaborate reasoning and inspection of requirements to create a stream processing system for heterogeneous data set. The stream processing paradigm requires solution which can provide  low latency, high throughput, fault tolerance along with scalability and versatility.  The system requires extensibility to plug and play different components to provide analytics for in-stream processing and stored stream data in a persistent storage. 


\section{Acknowledgements} 
I sincerely thank the reviewers for their comments and suggestions. This survey paper has been supported by the University of Helsinki as a learning initiative under Seminar course on distributed computing frameworks for Big Data.

%
\bibliographystyle{abbrv}
\bibliography{main}  
%
%

\balancecolumns
\end{document}